\documentclass[11pt]{article}

\usepackage{amsmath}
\usepackage{graphicx}
\usepackage{amsfonts}
\usepackage{amssymb}
\usepackage{epsfig}
\usepackage{color}
\usepackage{psfrag}
\usepackage{epstopdf}

\usepackage{caption}
\usepackage{subcaption}

\newcommand{\EQ}{\begin{equation}}
\newcommand{\EN}{\end{equation}}
\newcommand{\be}{\begin{equation}}
\newcommand{\ee}{\end{equation}}
\newcommand{\bea}{\begin{eqnarray}}
\newcommand{\eea}{\end{eqnarray}}

\begin{document} \setcounter{page}{0}
\newpage
\setcounter{page}{0}
\renewcommand{\thefootnote}{\arabic{footnote}}
\newpage
\begin{titlepage}
\begin{flushright}
\end{flushright}
\vspace{0.5cm}
\begin{center}
{\large {\bf Mass of quantum topological excitations\\ and order parameter finite size dependence}}\\
\vspace{1.8cm}
{\large Gesualdo Delfino$^{1,2}$ and Marianna Sorba$^{1,2}$}\\
\vspace{0.5cm}
{\em $^1$SISSA -- Via Bonomea 265, 34136 Trieste, Italy}\\
{\em $^2$INFN sezione di Trieste, 34100 Trieste, Italy}\\
\end{center}
\vspace{1.2cm}

\renewcommand{\thefootnote}{\arabic{footnote}}
\setcounter{footnote}{0}

\begin{abstract}
\noindent
We consider the spontaneously broken regime of the $O(n)$ vector model in $d=n+1$ space-time dimensions, with boundary conditions enforcing the presence of a topological defect line. Comparing theory and finite size dependence of one-point functions observed in recent numerical simulations we argue that the mass of the underlying topological quantum particle becomes infinite when $d\geq 4$. 
\end{abstract}

\end{titlepage}

\newpage
\section{Introduction}
Some quantum field theories allow for a nontrivial mapping between the ground state manifold and the spatial boundary, and then for topological excitations (see e.g. \cite{Coleman}). These excitations correspond to extended configurations of the fields entering the action, a feature which requires nonperturbative methods for their characterization as quantum particles. It is well known that these methods have been available in the case of space-time dimension $d=2$, as illustrated by sine-Gordon solitons: on one hand fermionization maps them onto the fundamental fields of the massive Thirring model \cite{Coleman_SG,Mandelstam}, on the other integrability provides the exact soliton scattering amplitudes \cite{ZZ}. 

In the last years it has been pointed out that the correspondence -- through analytic continuation to imaginary time -- between relativistic and Euclidean field theories can be exploited to gain insight into the case $d>2$ \cite{vortex,vortex_mass}. For this purpose one works in the spontaneously broken phase of the Euclidean theory, with boundary conditions enforcing the presence of a topological defect, and with a finite size $R$ in the imaginary time direction. Then the large $R$ asymptotics of one-point functions such as the order parameter are determined by the state with a single topological particle, and can be obtained analytically \cite{vortex}. In addition, comparison between these analytical results and their determination in numerical simulations of the corresponding spin system allows the measurement of basic parameters of the theory such as the mass of the topological particle; this program was illustrated in \cite{vortex_mass} for the case of the scalar $O(2)$ theory in $d=3$, which describes the universality class of the superfluid transition (see \cite{Lipa}) and possesses quantum vortex excitations.

More recently, the program of \cite{vortex_mass} has been carried through in \cite{PS1,PS2} for the $O(3)$ scalar theory in $d=4$. Intriguingly, the numerical simulations showed a scaling dependence on the parameters -- the finite size $R$ and the deviation from criticality -- markedly different from that observed in \cite{vortex_mass}. In this paper we show that the theory of \cite{vortex} accounts for both cases, with the difference arising from the fact that the mass of the topological particle is finite in the three-dimensional $O(2)$ model and infinite in the four-dimensional $O(3)$ model. We argue that this is due to the passage from the nontrivial renormalization group fixed point of the first case to the Gaussian fixed point of the second. 

The paper is organized as follows. In the next section we recall the theoretical setting before applying it to the order parameter in section~3 and to the energy density in section~4. Fluctuations for the case of infinite mass of the topological particles are discussed in section~5, while conclusive remarks are given in the last section.

\section{General setting}
We consider the universality class of $O(n)$-symmetric ferromagnets,  whose simplest representative (see e.g. \cite{Cardy}) is the vector model defined by the reduced Hamiltonian
\EQ
{\cal H}=-\frac{1}{T}\sum_{<i,j>}{\bf s}_i\cdot{\bf s}_j\,,
\label{H}
\EN
where $T$ is the temperature, ${\bf s}_i$ is a $n$-component unit vector located at site $i$ of a regular lattice, and the sum is performed over all pairs of nearest neighboring sites. Denoting by $T_c$ the critical temperature, we focus on the regime $T<T_c$ in dimension 
\EQ
d=n+1\geq 2\,.
\EN
Then the $O(n)$ symmetry of the Hamiltonian is spontaneously broken, i.e. $\langle{\bf s}_i\rangle\neq 0$, with $\langle\cdots\rangle$ denoting the average over spin configurations weighted by $e^{-{\cal H}}$. 

Close to $T_c$, where the intrinsic length scale becomes much larger than lattice spacing, the system is described by an $O(n)$-invariant Euclidean scalar field theory, which in turn is the continuation to imaginary time of a quantum field theory in $n$ space and one time dimensions. Switching to notations of the continuum, we denote by $({\bf x},y)$ a point in  Euclidean space, with $y$ the imaginary time and ${\bf x}=(x_1,\ldots,x_n)$, and by ${\bf s}({\bf x},y)$ the order parameter field, namely the continuous version of the lattice spin variable ${\bf s}_i$. Then the Landau-Ginzburg field theory takes the usual form specified by the action
\EQ
{\cal A}=\int d^dx\left\{[\partial_\mu{\bf s}(x)]^2+g_2\,{\bf s}^2(x)+g_4[{\bf s}^2(x)]^2\right\}\,,
\label{action}
\EN
with the $O(n)$ critical point reachable tuning the couplings (see e.g. \cite{Cardy}).

\begin{figure}[t]
    \centering
        \includegraphics[width=7cm]{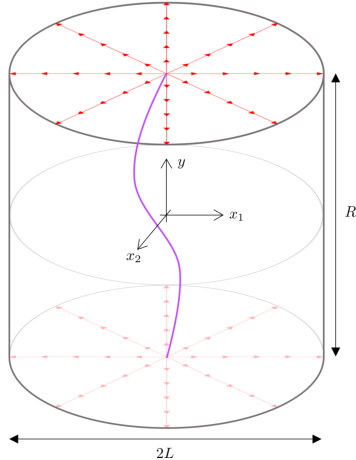}
    \caption{Geometry considered in the text ($n=2$), where it is understood that $L\to\infty$. The spins on the top and bottom surfaces are fixed to point radially outwards, so that a topological defect line (one configuration is shown) runs between the central points of these surfaces.}
    \label{geometry}
\end{figure}

Since the ground state manifold and the space boundary both correspond to the sphere $S^{n-1}$, 
the quantum theory possesses particle excitations associated with extended field configurations, with different points on the space boundary mapped onto different ground states. Such topological excitations are kinks in the 2D Ising model ($n=1$), vortices in the 3D XY model ($n=2$), hedgehogs in the 4D Heisenberg model ($n=3$), and so on. The propagation of these particles in imaginary time generates topological defect lines for the Euclidean system.

 We consider the system in the hypercylinder $|{\bf x}|\leq L$, $|y|\leq R/2$, with $L\to\infty$ and $R$ large but finite. The boundary conditions are chosen in such a way that the spin field ${\bf s}({\bf x},y)$ points outwards in the radial direction ${\bf x}/|{\bf x}|\equiv\hat{\bf x}$ on the hypersurfaces $|{\bf x}|=L,\,\,|y|<R/2$, and $0<|{\bf x}|\leq L,\,\,y=\pm R/2$. This leads to the formation of a topological defect on each section with constant $y$, with the defect center spanning as $y$ varies a line (particle trajectory) running between the endpoints at ${\bf x}=0$, $y=\pm R/2$.  The system geometry and boundary conditions\footnote{As long as the topology is preserved and the $L\to\infty$ limit is considered,  the system geometry does not need to be cylindrical for comparison with our subsequent analytical results in the continuum limit; see \cite{vortex_mass} for a parallelepiped realization which is equivalent for our purposes. } are illustrated in fig.~\ref{geometry} for the case $n=2$.
 
 It was shown in \cite{vortex} how the large $R$ asymptotics of one-point functions can be analytically determined in the theory specified above. We now recall the main points of that analysis. The boundary conditions at $y=\pm R/2$ act as boundary states $|B(\pm R/2)\rangle=e^{\pm\frac{R}{2}H}|B(0)\rangle$ of the Euclidean time evolution, where $H$ denotes the Hamiltonian of the quantum system. These boundary states can be expanded on the basis of asymptotic particle states of the quantum field theory\footnote{We refer here to the bulk theory, namely the fully translation invariant theory.  More generally, see e.g.  \cite{Ryder} for the basic formalism of quantum field theory.}, and will contain the topological particle $\tau$ as the contribution with minimal energy, namely
\EQ
|B(\pm R/2)\rangle=\int\frac{d{\bf p}}{(2\pi)^n E_{\bf p}}\,a_{\bf p}\,e^{\pm\frac{R}{2}E_{\bf p}}\,|\tau({\bf p})\rangle+\cdots\,,
\label{B}
\EN
where ${\bf p}$ is the $n$-component momentum of the particle, $E_{\bf p}=\sqrt{{\bf p}^2+m_\tau^2}$ its energy, $m_\tau$ its mass, $a_{\bf p}$ an amplitude, and we normalize the states by $\langle \tau({\bf p}')|\tau({\bf p})\rangle=(2\pi)^n E_{\bf p}\,\delta({\bf p}-{\bf p}')$. In the calculations performed with the boundary conditions we have chosen (which we will indicate with a subscript ${\cal B}$) the contribution in (\ref{B}) with one topological particle determines the asymptotics for $m_\tau R\gg 1$. In the following, the symbol $\sim$ will indicate omission of terms subleading in the large $R$ limit. To begin with we have 
\bea
Z_{\cal B} & \equiv & \langle B(R/2)|B(-R/2)\rangle=\langle B(0)|e^{-RH}|B(0)\rangle\nonumber\\
&\sim & |a_0|^2\int\frac{d{\bf p}}{(2\pi)^n m_\tau}\,e^{-(m_\tau+\frac{{\bf p}^2}{2m_\tau})R}=\frac{|a_0|^2}{m_\tau}\left(\frac{m_\tau}{2\pi R}\right)^{n/2}e^{-m_\tau R}\,.
\label{Z}
\eea
Similarly, the expectation value of a scalar field $\Phi$ is given by\footnote{We consider for simplicity $y=0$, the extension to $y$ generic being straightforward (see \cite{vortex_mass}).}
\bea
\langle\Phi({\bf x},0)\rangle_{\cal B}&=&\frac{1}{Z_{\cal B}}\,\langle B(R/2)|\Phi({\bf x},0)|B(-R/2)\rangle \nonumber\\
&\sim &\left(\frac{2\pi R}{m_\tau}\right)^{n/2}\int\frac{d{\bf p}_1d{\bf p}_2}{(2\pi)^{2n}m_\tau}\,F_\Phi({\bf p}_1|{\bf p}_2)\,e^{-\frac{R}{4m_\tau}({\bf p}_1^2+{\bf p}_2^2)+i{\bf x}\cdot({\bf p}_1-{\bf p}_2)},
\label{vPhi0}
\eea
where
\EQ
F_\Phi({\bf p}_1|{\bf p}_2)=\langle \tau({\bf p}_1)|\Phi(0,0)|\tau({\bf p}_2)\rangle\,,\hspace{1cm}{\bf p}_1,{\bf p}_2\to 0
\label{ff}
\EN
is the matrix element on the topological particle state, evaluated in the low-energy limit enforced by the large $R$ expansion. It decomposes as 
\EQ
F_\Phi({\bf p}_1|{\bf p}_2)=F^c_\Phi({\bf p}_1|{\bf p}_2)+(2\pi)^nE_{{\bf p}_1}\delta({\bf p}_1-{\bf p}_2)\,\langle\Phi\rangle\,,
\label{connected}
\EN
where $\langle\Phi\rangle$ is the bulk expectation value, and we see that only the connected part $F^c_\Phi$ contributes to the ${\bf x}$-dependence of (\ref{vPhi0}). If $F^c_\Phi$ behaves for small momenta as momentum to the power $\alpha_\Phi$,  rescaling of momentum components by $\sqrt{R}$ shows that the  ${\bf x}$-dependent part of (\ref{vPhi0}) is suppressed at large $R$ as 
\EQ
R^{-(n+\alpha_\Phi)/2}\,.
\label{suppression}
\EN

\section{Order parameter}
The order parameter $\langle{\bf s}({\bf x},0)\rangle_{\cal B}$ is an odd function of ${\bf x}$ which interpolates between zero at ${\bf x}=0$ and the asymptotic value
\EQ
\lim_{|{\bf x}|\to\infty}\langle{\bf s}({\bf x},0)\rangle_{\cal B}\sim v\,\hat{\bf x}\,,
\label{c1}
\EN
where 
\EQ 
v=|\langle{\bf s}({\bf x},y)\rangle|
\label{v}
\EN
is the modulus of the bulk magnetization. This interpolation is not suppressed as $R\to\infty$ and requires $\alpha_{\bf s}=-n$, and it was seen in \cite{vortex} that $F^c_{\bf s}({\bf p}_1|{\bf p}_2)$ is proportional to 
\EQ
\frac{{\bf p}_1-{\bf p}_2}{|{\bf p}_1-{\bf p}_2|^{n+1}}\,.
\label{ffs}
\EN
Upon insertion in (\ref{vPhi0}) this leads to \cite{vortex}
\EQ
\langle{\bf s}({\bf x},0)\rangle_{\cal B}\sim v\,\frac{\Gamma\left(\frac{n+1}{2}\right)}{\Gamma\left(1+\frac{n}{2}\right)}\,{}_1F_1\left(\frac{1}{2},1+\frac{n}{2};-z^2\right)z\,\hat{\bf x}\,,
\label{result}
\EN
where ${}_1F_1(\alpha,\gamma;z)$ is the confluent hypergeometric function, and
\EQ
z\equiv \sqrt{\frac{2m_\tau}{R}}\,|{\bf x}|\,.
\label{zeta}
\EN

For $n=1$ the result (\ref{ffs}) is the low energy limit of the matrix element known exactly \cite{BKW} from 2D Ising field theory, which is integrable (see \cite{review} for a review). On the other hand, (\ref{result}) reduces to $v\,\mbox{erf}(z)$; this result, which describes the separation of phases in the 2D Ising model, was obtained from the exact lattice solution in \cite{Abraham,Abraham_review} and from field theory in \cite{DV,DS} (see \cite{DSS_ising3D,DSS_wall} for the relation with phase separation in $d=3$). 

For $n=2$ the result (\ref{result}) was successfully tested against Monte Carlo simulations of the 3D XY model in \cite{vortex_mass}. In particular, this allowed to numerically determine the mass $m_\tau$ of the vortex particle, which was the only unknown parameter involved in the simulations. This finding is particularly relevant in view of Derrick's theorem \cite{Derrick} (see also \cite{Coleman}), which prevents the existence of finite energy topological configurations in theories of ${\it classical}$ self-interacting scalar fields in $d>2$. The finite value of $m_\tau$ measured in \cite{vortex_mass} provided the first direct verification that this obstruction does not in general persist at the quantum level. In particular, a result of classical field theory such as Derrick's theorem has no special reason to hold in presence of the nontrivial fixed point of the renormalization group exhibited by the 3D XY model.  

At the same time, the last observation suggests that something might change for $n\geq 3$. Indeed, $d=4$ is the upper critical dimension $d_c$ of the theory (\ref{action}), meaning that for $d\geq d_c$ the fixed point ruling the critical behavior is the Gaussian one, the role of fluctuations is suppressed and the critical exponents take mean field values (see e.g. \cite{Cardy}). Derrick's result might persist in this case and it is relevant to see what the above analysis predicts for $m_\tau\to\infty$. In this case, for any finite $R$, (\ref{zeta}) yields $z\to\infty$ as long as ${\bf x}\neq 0$, and the result (\ref{result}) for the order parameter becomes
\EQ
\lim_{m_\tau\to\infty}\langle{\bf s}({\bf x},0)\rangle_{\cal B}\sim\left\{
\begin{array}{l}
v\,\hat{\bf x}\,,\hspace{1cm}{\bf x}\neq 0\,,\\
\\
0\,,\hspace{1.2cm}{\bf x}=0\,.
\end{array}
\right.
\label{result2}
\EN
It follows that, if the topological particle has an infinite mass, the order parameter becomes $R$-independent in the large $R$ limit we consider. The absence of an appreciable $R$-dependence of the one-point functions is the key difference observed in the numerical simulations of \cite{PS1,PS2} for $n=3$ with respect to those of \cite{vortex_mass} for $n=2$. We now see that this difference is explained by the theory and indicates that the topological mass $m_\tau$ is infinite for $n=3$, i.e. for $d=4$. The same is then expected to hold more generally for $d\geq d_c=4$, namely in presence of a Gaussian critical point. Spontaneous symmetry breaking around a Gaussian point is taken into account already at the classical level, and $m_\tau=\infty$ means that Derrick's result of classical field theory persists in the mean field regime. It is worth stressing how $m_\tau=\infty$ does not mean that the topological particle is absent: the result (\ref{result2}) is entirely due to this particle.  In other words,  the infinitely massive particle does not contribute to fluctuations but provides the topological charge required when the boundary conditions enforce the presence of a topological defect.

\section{Energy density}
\begin{figure}[t]
    \centering
    \begin{subfigure}[h]{0.45\textwidth}
        \includegraphics[width=\textwidth]{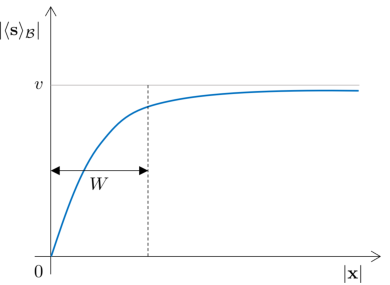}
    \end{subfigure}\hspace{1cm}%
    \begin{subfigure}[h]{0.45\textwidth}
        \includegraphics[width=\textwidth]{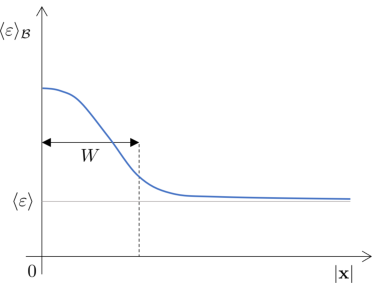}
    \end{subfigure}
    \caption{Qualitative profiles at $y=0$ for the modulus of the order parameter (left) and the energy density (right) as a function of the distance from the center. The width $W$ (eq.~(\ref{W})) of the pre-asymptotic region is replaced by $\tilde{W}$ (eq.~(\ref{Wtilde})) when the mass of the topological particle becomes infinite ($n\geq 3$). }
    \label{profiles}
\end{figure}

It is interesting to extend the analyis to the energy density field $\varepsilon\propto{\bf s}^2$, which was also simulated in \cite{vortex_mass,PS1}. Recalling (\ref{suppression}) and (\ref{zeta}), the result of (\ref{vPhi0}) for this field will take the form
\EQ
\langle\varepsilon({\bf x},0)\rangle_{\cal B}\sim \left[\frac{f_\varepsilon(z)}{(m_\tau R)^{(n+\alpha_\varepsilon)/2}}+1\right] \langle\varepsilon\rangle\,,
\label{epsilon}
\EN 
where $f_\varepsilon$ depends on the specific form of the connected matrix element $F^c_\varepsilon({\bf p}_1|{\bf p}_2)$ for small momenta. It follows from (\ref{vPhi0}) and (\ref{connected}) that the $|{\bf x}|$-dependent term in (\ref{epsilon}) is the contribution to the energy density on the hyperplane $y=0$ coming from the propagation of the topological particle between the endpoints $({\bf x},y)=(0,\pm R/2)$ of its trajectories. Hence, the dimensionless function $f_\varepsilon(z)$ is proportional to the probability of finding the particle at a distance $|{\bf x}|$ from the origin on that hyperplane, and monotonically decreases from $f_\varepsilon(0)$ to $f_\varepsilon(\infty)=0$; the limit
\EQ 
\lim_{|{\bf x}|\to\infty}\langle\varepsilon({\bf x},0)\rangle_{\cal B}\sim\langle\varepsilon\rangle\,,
\label{eps}
\EN 
with $\langle\varepsilon\rangle$ the bulk energy density, is the expected one. The form (\ref{zeta}) of the scaling variable $z$ shows that the width $W$ of the peak of (\ref{epsilon}) around ${\bf x}=0$ (fig.~\ref{profiles}) depends on the parameters as
\EQ
W\propto\sqrt{R/m_\tau}\propto\sqrt{(T_c-T)^{-\nu}R}\,,
\label{W}
\EN
where $\nu$ is the correlation length critical exponent. Hence, (\ref{epsilon}) becomes flat as $R\to\infty$, and (\ref{eps}) requires $\alpha_\varepsilon>-n$. The dependence (\ref{epsilon}) of the energy density for large $R$ is known in full detail for $n=1$ \cite{DV,DS,ST}, and has been confirmed numerically for $n=2$ \cite{vortex_mass}. Passing to the case $n\geq 3$, we know by now that it requires the limit $m_\tau\to\infty$. Knowing that $f_\varepsilon(\infty)=0$ and $\alpha_\varepsilon>-n$, (\ref{epsilon}) yields
\EQ 
\lim_{m_\tau\to\infty}\langle\varepsilon({\bf x},0)\rangle_{\cal B}\sim\langle\varepsilon\rangle\,.
\label{eps2}
\EN 
This result explains, in particular, why no appreciable $R$-dependence of the energy density was observed in the simulations of \cite{PS1} for $n=3$. 

\section{Residual fluctuations}
An additional element which complicated the interpretation of the numerical results of \cite{PS1,PS2} for $n=3$ is that, in spite of the $R$-independence that we have now explained, the overall qualitative ${\bf x}$-dependence of the one-point functions was found to be quite analogous to that observed in \cite{vortex_mass} for $n=2$. In particular, $\langle{\bf s}({\bf x},0)\rangle_{\cal B}$ was found to exhibit a smooth interpolation between zero at ${\bf x}=0$ and $v\,\hat{\bf x}$ at $|{\bf x}|=\infty$, at variance with the step-like interpolation of (\ref{result2}). The energy density $\langle\varepsilon({\bf x},0)\rangle_{\cal B}$ was observed to display a bell shape centered in ${\bf x}=0$ and approaching the bulk value $\langle\varepsilon\rangle$ for $|{\bf x}|$ large enough. 

These corrections to (\ref{result2}) and (\ref{eps2}) should come from contributions not considered in the previous discussion. In the Ising case ($n=1$) it is known \cite{DV,DS} that the leading corrections to (\ref{result}) and (\ref{epsilon}) expand in powers of $R^{-1/2}$ and are due to the subleading terms of the expansion for small momenta associated to the state $|\tau\rangle$ itself\footnote{See \cite{ST} for an accurate comparison between theoretical predictions and the results of numerical simulations.} in (\ref{vPhi0}). For $n\geq 3$, however, this type of corrections are eliminated by the divergence of $m_\tau$, and we should consider the states contributing to the dots in (\ref{B}). These are of the type $\tau$ (which provides the required topological charge) plus Goldstone bosons associated to the spontaneous breaking of the continuous symmetry. The analytical evaluation of the contribution of these states to the one-point functions would require information about the matrix elements of the fields on these states, which is not available.  Remarkably,  however,  we now show that implications sufficient for our purposes can be obtained from the following considerations.  For $n\geq 3$ the nonzero width $\tilde{W}$ of the pre-asymptotic region in the profiles of fig.~\ref{profiles} -- i.e.  the deviation from the results (\ref{result2}) and (\ref{eps2}) -- is due to the Goldstone fluctuations.  Since $m_\tau=\infty$ suppresses the $R$-dependence\footnote{It cannot be excluded that the cumulative effect of Goldstone bosons results in a residual, very weak -- e.g. logarithmic -- $R$-dependence which was not detected within the numerical accuracy of the simulations in \cite{PS1,PS2}. For the purpose of explaining the data of \cite{PS1,PS2}, this possibility can be consistently ignored.}, the width $\tilde{W}$ can only depend on the temperature and scales in the way expected for a length,
\EQ
\tilde{W}\propto (T_c-T)^{-\nu}\,,
\label{Wtilde}
\EN
where the critical exponent $\nu$ takes the mean field value $1/2$ around the Gaussian fixed point relevant for $n\geq 3$. If one tries to explain the scaling observed in simulations performed for $n\geq 3$ through the formulae which apply to the case of $m_\tau$ finite ($n=1,2$), this means reproducing the behavior (\ref{Wtilde}) using (\ref{W}),  namely writing $(T_c-T)^{-\nu}\propto\sqrt{R/m_\tau}$.  One is then led to the {\it formal} identification $m_\tau\propto (T_c-T)^{2\nu}R=(T_c-T)R$.  This is precisely what was observed using (\ref{result}) for the fits of \cite{PS1,PS2} at $n=3$. We now see why the $R$-dependence of $m_\tau$ obtained in this way is artificial and, at the same time, how the data of \cite{PS1,PS2} confirm $m_\tau=\infty$ and (\ref{Wtilde}).

\section{Conclusion}
In this paper we considered $d$-dimensional statistical models in their spontaneously broken phase, with boundary conditions enforcing the presence of a topological defect line. Within the correspondence with quantum field theory in $d-1$ spatial dimensions, the defect line corresponds to the trajectory of a topological particle propagating for a large but finite imaginary time $R$. To be specific we referred to the case of the $O(n)$ vector model in $d=n+1$ dimensions, for which the presence of quantum topological excitations follows from the fact that the ground state manifold and the spatial boundary both correspond to the hypersphere $S^{n-1}$. Recent Monte Carlo simulations for the cases $n=2$ \cite{vortex_mass} and $n=3$ \cite{PS1,PS2} showed different scaling dependence of one-point functions (e.g. the order parameter) on the parameters of the theory, namely the finite size $R$ and the deviation from critical temperature. We showed in this paper that the theory of \cite{vortex} accounts for both cases, the difference being produced by a mass $m_\tau$ of the topological particle which is finite for $n=2$ and infinite for $n=3$. We argued that this is due to the fact that $d=4$ is the upper critical dimension of the $O(n)$ model. For $d\geq 4$ the critical behavior is controlled by the Gaussian fixed point, namely the fixed point explicitly present in the Landau-Ginzburg action (\ref{action}). This action belongs to the class covered by Derrick's theorem \cite{Derrick,Coleman} of {\it classical} field theory, which states that static solutions in self-interacting scalar theories in $d>2$ have infinite energy. The Monte Carlo data of \cite{PS1,PS2} and their theoretical interpretation of the present paper indicate that Derrick's result gets through to the mean field regime $d\geq 4$, in the sense that topological particles have infinite mass. For $d<4$, instead, the critical behavior is controlled by a nontrivial fixed point, for which arguments of classical field theory have no reason to remain quantitatively reliable. In particular, the $R$-dependence of one-point functions observed numerically in \cite{vortex_mass} for $n=2$ showed that the quantum vortex has a finite mass which was estimated from the comparison between theory and Monte Carlo data.  It is worth recalling that this is a particularly relevant result in view of the long debate concerning the definition of a mass of vortices in superfluids (see \cite{vortex_mass} and references therein),  a debate in which the transposition of considerations of classical field theory (Derrick's theorem) to the quantum case plays a substantial role.  The analysis of our present paper gives concrete evidence that such a transposition is possible only in the mean field regime $d\geq d_c$.  It is remarkable that this insight could be obtained comparing considerations of quantum field theory with numerical simulations performed in the Euclidean case,  thus providing a very fruitful operative illustration of the interplay between real and imaginary time.  In perspective,  it would be very interesting to numerically test our predictions in the $O(n)$ model  for $n>3$,  as well as for other symmetries.


\end{document}